\documentclass[%
 preprint,
 amsmath,amssymb,
 aps,
]{revtex4-2}

\usepackage{graphicx}
\usepackage{dcolumn}
\usepackage{bm}
\usepackage{hyperref}
\usepackage{color}
\usepackage{booktabs}

\begin{document}


\title{ Influence of the Pauli exclusion principle on the $\alpha$ decay  \\}

 \author{M. M. Amiri\,}%
 \thanks{Email: morteza.moghaddari@stu.umz.ac.ir}%
 \author{O. N. Ghodsi}%

 \affiliation{Department of Physics, Faculty of Science, University of Mazandaran, P.O. Box 47415-416, Babolsar, Iran.}%

\date{\today} %

\begin{abstract}

In this study, the effects of repulsive nucleon-nucleon interactions arising from the Pauli exclusion principle were examined regarding the half-lives of heavy even-even nuclei with $84\leq Z\leq92$. The Pauli exclusion principle is applied to our investigations by renormalizing the nucleon-nucleon interactions according to the Bohr-Sommerfeld quantization condition. It is also applied to the double-folding ($DF$) formalism as a modification term by investigating the kinetic energy variation at the overlapping regions between the densities of the alpha and daughter nuclei. The standard deviation for the calculated half-lives from their corresponded experimental data is 0.340 for the renormalized DF formalism. Its value reduces to 0.309 for the DF formalism modified with the kinetic energy contribution estimated by the extended Thomas-Fermi approach. The calculated $\alpha$ decay half-lives are more consistent with experimental data if the kinetic energy contribution is being associated with the DF formalism.
\end{abstract}

\maketitle


\section{\label{sec:level1} Introduction}

Over the past decades, the $\alpha$ decay process has attracted much interest toward investigating the decay and fusion mechanisms of heavy and superheavy nuclei \cite{1,711,85,90,91,53,54,2,3,4,5,6,7}. Various approaches have been adopted to investigate the properties of the $\alpha$ decay processes \cite{8,9,10,11,12,13}. These approaches are based on nucleon-nucleon interactions or density functionals of the emitted cluster and daughter nuclei \cite{14,15,16}.

The properties of cluster and daughter nuclei play a significant role in calculating the interaction potential between two interacting nuclei. The nucleon density distribution is one of the most critical factors in assessing nuclear properties that can affect the interaction potential \cite{17,18,19,20,21} and thereby confront theoretical calculations with experimental data. Meanwhile, investigations of cluster decay and nuclear fusion are predominantly relevant for the interaction potentials in the partial and full overlap density regions \cite{22,23,24}. Although the semi-microscopic DF potentials based on the effective M3Y nucleon-nucleon ($NN$) interaction can well reproduce most of the scattering data, they fail in describing many reactions that are strongly affected by the characteristics of the potential below the barrier in the internal region \cite{25}. This deficiency can be due to the non-consideration of the repulsive core in the DF formalism \cite{26}. However, the Pauli exclusion principle ($PEP$) satisfactory and antisymmetrization impact would be important ones compensating such insufficiencies \cite{27}. To this goal, the contribution of an increase in the intrinsic kinetic energy of the nuclear densities at small separation distances can be employed due to the antisymmetrization of both the interaction matrix element and the distortion \cite{28}.

As a result of the repulsive force would be remarkable at minimal distances, the internal kinetic energy reaches a maximum value when the densities of participating nuclei have a complete overlap. On the other hand, the PEP antisymmetrization and nuclear incompressibility would be an obstacle to the occurrence of a complete overlap in the total system that guarantees saturation properties of nuclear matter \cite{29}.

In some formalism like DF, the sudden approximation is being used. The densities of the colliding nuclei are assumed to be unchanged at all distances during the overlapping process. The densities of two interacting nuclei overlap so that an increasing repulsive force would be expected due to this approximation and PEP. In some theoretical studies, this increase in energy of the dinuclear system is being simulated within a repulsive force in the NN interaction, and the nucleus-nucleus potential is being modified in consequence. Such modifications have been made mostly for fusion reactions, which led to the better reproduction of experimental data. For instance, it is shown that the incorporation of Pauli blocking properly with the DF model can achieve by taking into account the redefinition of the density matrices of the free isolated nuclei \cite{27}. Microscopically, from the standard M3Y potential developed with the density-dependent Pauli blocking potential of the density overlap of two colliding nuclei and consequent appearance some shallow pockets in $\alpha$-core potentials of $\alpha$+$^{208}$Pb, $\alpha$+$^{197}$Au, $\alpha$+$^{209}$Bi, and $\alpha$+$^{238}$U, their originated fusion cross-sections were consistent with the experimental data \cite{57}. Also, the simultaneous cross-section studies based on the experimental evidence for the reactions $^{12}$C + $^{12}$C, $^{12}$C + $^{16}$O, and $^{16}$O + $^{16}$O proposed a soft Gaussian repulsive core to the Woods-Saxon potential \cite{30}. Moreover, the unexpected steep falloff of fusion cross sections at energies far below the Coulomb barrier for the reactions $^{64}$Ni + $^{64}$Ni, $^{58}$Ni + $^{58}$Ni, and $^{64}$Ni + $^{100}$Mo were successfully explained by combination using the DF potential based on the M3Y interactions supplemented with a simulated repulsive core \cite{24,29}.

On the other hand, such an additional repulsive force increases the kinetic energy of the total interaction \cite{32,57}. Therefore, the desired modifications can be investigated by estimating the variation of the kinetic energy in the dinuclear system. An increase in kinetic energy acts as a repulsive force in the dinuclear system that prevents the unexpected increase of the density overlapping at interior regions. Despite the well-illustrated overlapping effect in the density functional theory ($DFT$) \cite{59,60}, the effect of Pauli blocking is not well embedded in the DF formalism. For achieving such prevention, a repulsive force can be simulated in the DF formalism due to the PEP, which plays a similar role as like as the kinetic energy contribution in the DFT. Therefore, incorporating the PEP with the DF would be associated with a repulsive core simulation, which is one of the possible solutions to conserve the dinuclear system around the saturation density. To this aim, we intend to examine the influences of the PEP on the $\alpha$ decay within the DF formalism.

Furthermore, the $\alpha$ decay process is also a low-energy phenomenon, and it can not actually cause a sensible variation in the ground-state properties of an alpha emitter\cite{9,80}. On the other hand, it is shown that an $\alpha$ cluster is mostly formed in the pre-surface region of the nucleus. Some sensible density overlapping between alpha and daughter nuclei in more dense regions due to the considering the effect of Pauli blocking from the saturated core density \cite{75}. Therefore, the interior regions of the interaction potentials between alpha and daughter nuclei would be affected by the PEP, where the tunnelling process occurs. One of the possible mechanism that is extensively being used for simultaneously applying the PEP and clusterization state in the dinuclear system is the renormalizing the strength of effective NN interaction due to the Bohr-Sommerfeld (BS) quantization condition \cite{40,59,69,70,71,711,85}. This procedure compensates for the deficiency of the PEP in the DF formalism to some extent.

Since the effect of the Pauli blocking behaves as a repulsive force in the dinuclear system\cite{32}, an increase in kinetic energy in the dinuclear system would be expected due to densities overlap. Due to the fact that the effect of Pauli blocking is not embedded in the DF formalism, some promotions would be required to properly incorporate the DF formalism and the effect of Pauli blocking for the $\alpha$ decay studies. On the other hand, the Hartree-Fock model based on Skyrme forces is a successful approach for examining the ground-state properties of nuclei. With a similar approach adopted in the Hartree-Fock model, and different from what the BS condition suggests, we intend to estimate an increase in the kinetic energy due to the Pauli blocking within the extended Thomas-Fermi approach. Consequently, the $\alpha$ decay half-lives of even-even heavy nuclei will be calculated that are being affected by embedding kinetic energy causing by folding the densities of alpha and daughter nuclei into the nucleon-nucleon M3Y interactions in the DF formalism. A comparison between these results with their corresponding values obtaining through the renormalizing the strength of effective NN interaction of total potential due to the BS condition will be presented.

This paper is organized as follows: The formalism of potential, half-life calculations in Sec.~\ref{sec.2}, and our result and discussion are given in Sec.~\ref{sec.3}. This paper ends with the main results and conclusions presented in Sec.~\ref{sec.4}.

\section{\label{sec.2}Theoretical framework}

\subsection{\label{sec:level2}Double-Folding formalism and the $\alpha$ decay half-life}

The total potential $V(R)$ is written by
\begin{equation}\label{Eq.1}
  V(R) = V_{C}(R)+  V_{N}(R),
\end{equation}
where $V_{C}$, $V_{N}$ are the coulomb and nuclear parts of the total potential, respectively. Also, $R$ denotes the vector joining the center of masses of the two nuclei. In this study, we investigate the $\alpha$ decay of even-even nuclei that the transferred angular momentum for these decay processes are zero. The nuclear part is obtained by the double-folding model within folding the densities of the alpha and the daughter nuclei with the effective M3Y interaction,
\begin{align}\label{Eq.3}
V_{N}(R)&=\lambda_{0} V_{F}(R)\notag \\
&= \lambda_{0} \iint{\rho_{1}(\textbf{r}_{1})\, V_{eff}(\textbf{s}) \, \rho_{2}(\textbf{r}_{2})\, d^{3}\textbf{r}_{1}\,d^{3}\textbf{r}_{2}},
\end{align}
where, $\textbf{s}=\textbf{R}+\textbf{r}_1-\textbf{r}_2$ corresponds to the distance between two specified interacting points of the interacting nuclei, whose radius vectors are $\textbf{r}_1$ and $\textbf{r}_2$, respectively. The $ V_{eff}(\textbf{s})$ is an effective nucleon-nucleon interaction \cite{26,34}. The energy-dependent M3Y Reid-NN forces with zero range approximation that used in our calculations have following explicit forms \cite{35}:
\begin{align}\label{Eq.4}
V_{eff}&(\textbf{s})= 7999\frac{\exp(-4s)}{4s}-2134\frac{\exp(-2.5s)}{2.5s}+J_{00}\delta(\textbf{s}) \notag \\
&J_{00}= -276(1-0.005E/A),
\end{align}
where $E$ and $A$ are the incident energy in the center-of-mass frame and the mass number of the alpha particle, respectively. In Eq.(\ref{Eq.3}), $\rho_{1}$ is taken for the density distribution function of the spherical $\alpha$ particle in its Gaussian form used in this study \cite{33} that is given as
\begin{equation}\label{Eq.5}
\rho(r)= 0.4229\,e^{-0.7024\,r^2},
\end{equation}
and $\rho_2$ is the density of the daughter nucleus that is determined by Hartree-Fock-Bogoliubov calculations based on the set of Skyrme SLy4 parametrization \cite{36} as result of its capability for well reproducing the $\alpha$ decay energies of heavy and superheavy nuclei, in this study \cite{41}.

The parameter $\lambda_{0}$ in Eq.(\ref{Eq.3}) changes the folded potential strength that is known as the strength parameter. It can be determined by using the BS quantization condition~\cite{64,65,66}:
\begin{equation}\label{Eq.6}
\int_{R_{1}}^{R_{2}}\sqrt{\frac{2\mu}{\hbar^{2}}|V(R)-Q|}\,dR= (2n + 1)\frac{\pi}{2}=(G-\ell+1)\frac{\pi}{2},
\end{equation}
where $R_{2}$, $R_{3}$ are classical turning points are obtained by $V(R) = Q$ (the $\alpha$ decay energy), and for $0^{+}\rightarrow 0^{+}$ $s$-wave decay the inner turning point is at $R_{1}=0$.

The global quantum number $G$ of a cluster state can be obtained by the Wildermuth condition~\cite{40}:
\begin{equation}\label{Eq.7}
G = 2N + \ell = \sum_{i=1}^{4}g_{i},
\end{equation}
where $N$ is the number of nodes of the $\alpha$-core wave function; $\ell$ is the relative angular momentum of the cluster motion, and, $g_{i}$; is the oscillator quantum number of a cluster nucleon. For the $\alpha$ decay, we can take $G$ as
\begin{equation}\label{Eq.8}
  G = 2N + \ell =\begin{cases}
    18, & \text{\qquad  $N\leq82$},\\
    20, & \text{\qquad  $82<N\leq126$},\\
    22. & \text{\qquad  $N>126$.}
  \end{cases}
\end{equation}

The half-life of the $\alpha$ decay is $T_{1/2}=\hbar \ln2/\Gamma_{\alpha}$. In this relation, $\Gamma_{\alpha}$ is the $\alpha$ decay width of the cluster state within the Gurvitz and K\"{a}lbermann method, determined as \cite{67}
\begin{equation}\label{Eq.9}
\Gamma_{\alpha}= F P_{\alpha}\frac{\hbar^{2}}{4\mu}\exp\left( -2\int_{R_{2}}^{R_{3}} k(R)\,dR \right),
\end{equation}
where $F$ is normalization factor can be defined as below
\begin{equation}\label{Eq.10}
F\int_{R_1}^{R_2}\frac{dR}{2k(R)}\,=1,
\end{equation}

The $k(R)=(2\mu/\hbar^2[V(R)-Q])^{1/2}$ is the wave number. Also, $P_{\alpha}$ and $\mu$ are the alpha formation probability and reduced mass, respectively.

\subsection{\label{sec:level2} Alpha decay half-life within Wentzel-Kramers-Brillouin approximation}

In the Preformed Cluster Model (PCM) viewpoint, a formation probability is being attributed to the alpha cluster before the tunneling process. Consequently, the $\alpha$ decay half-lives can be calculated by $T_{1/2}={\ln2}/{\lambda}$ with the PCM viewpoint, where $\lambda$ is decay constant that is a multiplication of the barrier penetrability P, the assault frequency $\nu_{0}$, and cluster formation probability $P_{\alpha}$. Moreover, many researchers adopted $P_{\alpha}=1$, fission-like model, to calculate the $\alpha$ decay half-lives. However, in this study, the calculation of the $\alpha$ decay half-lives will be done the same way for both conditions. It should be noted that the strength parameter is assumed to be $\lambda_{0}=1$ for calculating the $\alpha$ decay half-lives through the WKB approximation.

The barrier penetrability $P$ can be calculated by semi-classical Wentzel-Kramers-Brillouin (WKB) approximation
\begin{equation}\label{Eq.12}
P=\exp\left(-\frac{2}{\hbar}\int_{R_{4}}^{R_{5}}\sqrt{\frac{2\mu}{\hbar^2}|V(R)-Q_{\alpha}|} dR\right)\,.
\end{equation}
where $R_{4}$, $R_{5}$  are classical turning points. In this study, the experimental $Q_{\alpha}$-values are taken from refs. \cite{51,52}. By assuming that the $\alpha$ particle vibrates in a harmonic oscillator potential with oscillation frequency $\omega$, the assault frequency $\nu_{0}$ can be determined as illustrated in ref.\cite{38}.

\subsection{\label{sec:level2}Cluster Formation Model}
In the cluster formation model (CFM), it is assumed that the parent nucleus is a compilation of different cluster states \cite{9}. For each preformation, there is a different wave function and a different Hamiltonian. Therefore, we assume that for each preformation or clusterization, there is a clusterization state represented by a wave function. If the parent nucleus has $N$ different clusterization states with total energy $E$, the Hamiltonian $H_{i}$ belongs to the $i$th clusterization defined with an $i$th wave function, therefore
\begin{equation}\label{Eq.14}
H_{i}\Psi_{i}=E\Psi_{i} \qquad i=1,2,\dots,N.
\end{equation}

Therefore, this nucleus is described by a total time-independent wave function that is a linear combination of these clusterization orthonormalized wave functions
\begin{equation}\label{Eq.15}
\Psi=\sum^{N}_{i=1}a_{i}\Psi_{i},
\end{equation}
where $a_{i}$ are the amplitudes for the clusterization states of the complete set and within the orthogonality condition,
\begin{equation}\label{Eq.16}
\sum_{i}^{N}|a_{i}|^{2}=1.
\end{equation}

Each cluster has specific formation energy $E_{fi}$ that
\begin{equation}\label{Eq.17}
E_{fi}= |a_{i}|^{2}E.
\end{equation}

The probability of the alpha clusterization state $P_{\alpha}$ is equivalent to $a_{\alpha}^{2}$. It can be calculated as
\begin{equation}\label{Eq.18}
P_{\alpha}=|a_{\alpha}|^{2}=\frac{E_{f\alpha}}{E}.
\end{equation}
here, $a_{\alpha}$ and $E_{f\alpha}$ denote the coefficient of the $\alpha$ clusterization and the formation energy of an $\alpha$ cluster. $E$ is composed
of the $E_{f\alpha}$ and the interaction energy between $\alpha$ cluster and daughter nuclei. The detailed illustrations are provided in ref.\cite{9}. In the framework of CFM, the $\alpha$ cluster formation energy $E_{f\alpha}$ and total energy $E$ of a considered system can be expressed as
\begin{align}\label{Eq.19}
E_{f\alpha} =&3B(A,Z)+B(A-4,Z-2)\notag\\
&-2B(A-1,Z-1)-2B(A-1,Z),
\end{align}
\begin{equation}\label{Eq.20}
E = B(A,Z)-B(A-4,Z-2),
\end{equation}
$B(A, Z)$ is the binding energy of the nucleus with mass number $A$ and proton number $Z$. The defined energies in Eq.\ref{Eq.19} and Eq.\ref{Eq.20} belong to even-even nuclei, and for an odd atomic number or odd neutron number, the formation energies can be found in refs. \cite{45,72}.

Moreover, the formation probability of each cluster state calculates by the CFM can well reproduce a more realistic formation probability, which follows the calculation of Varga et al. \cite{73,74}.

\section{\label{sec.3}Results and discussions}

Generally, the order of magnitude of alpha tunneling is about $10^{-21}$ s \cite{8,38,42} that proceed with the $\alpha$ decay studies are being done under sudden approximation. Also, this approximation is being used in DF formalism, in which the densities of the interacting nuclei are being assumed to be frozen at all distances during the interaction. One can expect that the densities of the nucleons in the compound system become increased as the folding density distributions of alpha and daughter nucleus begins at the nuclear surface and rise to total overlap. Meanwhile, the PEP would be more apparent in following this accumulation of nucleons. Hence, a variation in the kinetic energy at constant volume would be expected \cite{27,32}. In this case, such kinetic energy is not attributed to the kinetic energy of the emitted alpha particle. Hence, the $\alpha$-core interaction potential that is being calculated by DF formalism can be modified by adding the mentioned kinetic energy term that causing by overlapping densities of the alpha and daughter nuclei.

For estimating the kinetic energy well illustrated in the DFT \cite{16,43}, the self-consistent Hartree-Fock calculations are being performed comprising SLy4 Skyrme interaction. The variation of the kinetic energy of the density overlap of two colliding nuclei based on the DFT is being obtained by
\begin{align}\label{Eq.21}
\Delta K(R)=&\frac{\hbar^{2}}{2m}\iint \bigg\{\tau[\rho_{1p}(\mathbf{r})+\rho_{2p}(\mathbf{r}-\mathbf{R}),\rho_{1n}(\mathbf{r}) \notag\\
&+\rho_{2n}(\mathbf{r}-\mathbf{R})] -\tau[ \rho_{1p}(\mathbf{r})+\rho_{1n}(\mathbf{r})]\notag\\
&- \tau[ \rho_{2p}(\mathbf{r})+\rho_{2n}(\mathbf{r})]\bigg\} d\mathbf{r}.
\end{align}
where $\tau$ denotes the kinetic energy density. The two nuclei are overlapping at $R$ and completely separated at infinity $R=\infty$. The contribution of the kinetic energy density for the dinuclear system coincide with overlapping the densities of the alpha and daughter nuclei would be clarified by the extended Thomas-Fermi approach (ETF) and considering the semi-classical correction of the second-order $\hbar$ \cite{44} proposed as
\begin{align}\label{Eq.22}
\tau_{q}(\textbf{r})=& \frac{3}{5}(3\pi^{2})^{\frac{2}{3}}\rho_{q}^{\frac{5}{3}}+\frac{1}{36}\frac{(\mathbf{\nabla} \rho_{q})^{2}}{\rho_{q}}+ \frac{1}{3}\Delta \rho_{q}+\frac{1}{6}\frac{\mathbf{\nabla} \rho_{q}.\mathbf{\nabla} f_{q}}{f_{q}} \notag\\
&+\frac{1}{6}\rho_{q}\frac{\Delta f_{q}}{f_{q}}-\frac{1}{12}\rho_{q}\left(\frac{\mathbf{\nabla} f_{q}}{f_{q}}\right)^{2} \notag \\
&+\frac{1}{2}\rho_{q}\left(\frac{2m}{\hbar^2}\right)^2 \left(\frac{W_{0}}{2}\frac{\mathbf{\nabla}(\rho+\rho_{q})}{f_{q}}\right)^2,
\end{align}
where $q$ denotes proton and neutron and $f_{q}(\textbf{r})$ is the effective mass form factor that is given as
\begin{align}\label{Eq.23}
f_{q}(\textbf{r})=& 1+ \frac{2m}{\hbar^{2}}\frac{1}{4}\left[t_{1}\left(1+\frac{x_{1}}{2}\right)+ t_{2}\left(1+\frac{x_{2}}{2}\right)\right]\rho(\textbf{r}) \notag \\
&-\frac{2m}{\hbar^{2}}\frac{1}{4}\left[t_{1}\left(x_{1}+\frac{1}{2}\right)- t_{2}\left(x_{2}+\frac{1}{2}\right)\right]\rho_{q}(\textbf{r}).
\end{align}
the parameters $x_1$, $x_2$, $t_1$, $t_2$, and $W_0$ are obtained by fitting different properties of nuclei, $m$ and $\rho=\rho_{1}+\rho_{2}$ are the nucleon mass and nuclear densities, respectively.

\begin{figure}
\centering
\includegraphics[width=4.0in,height=7.5in]{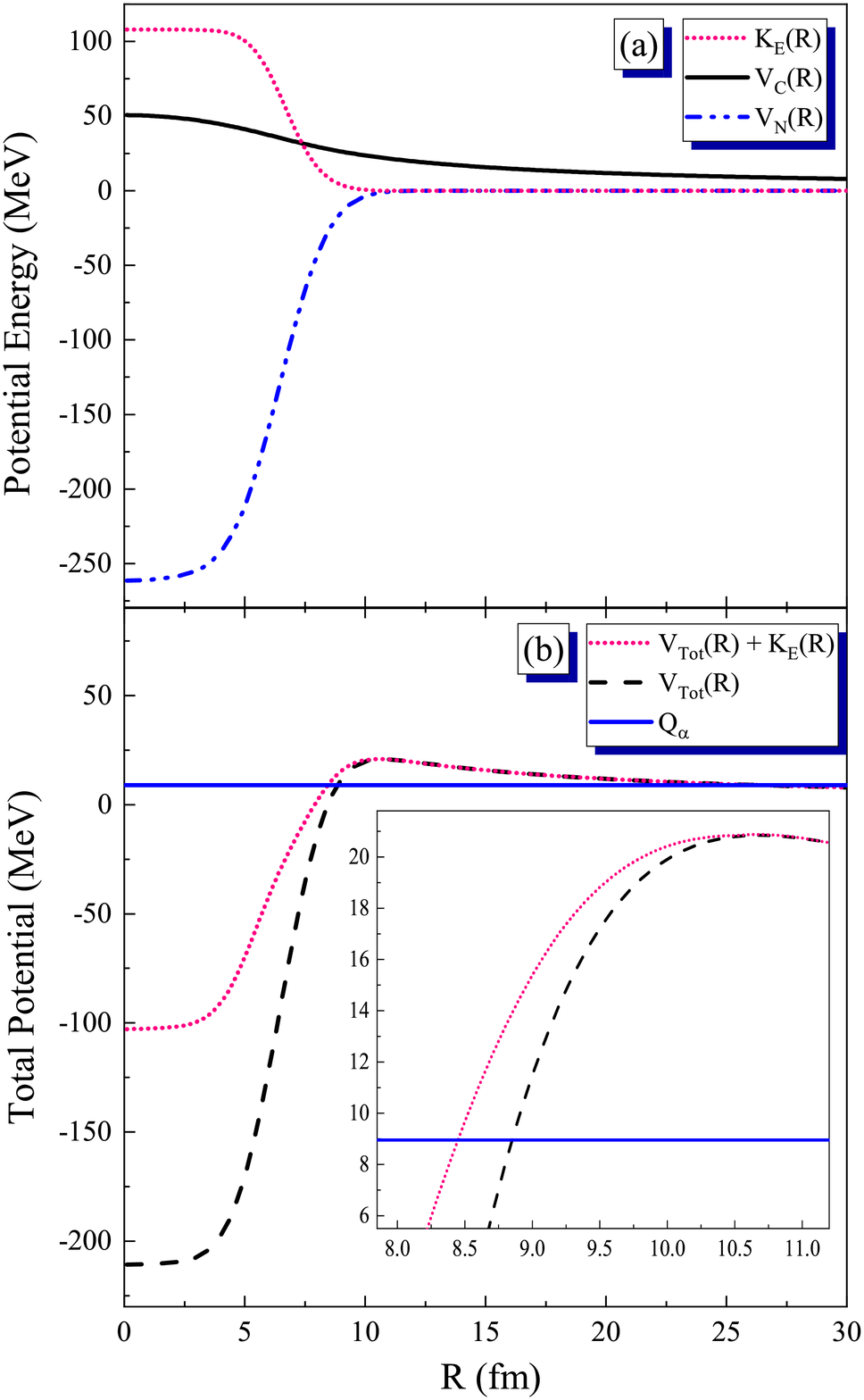}
\caption{\label{Fig02} (a) The calculated $V_{C}(R)$, $V_{N}(R)$, and kinetic energy, $K_{E}(R)$, typically for $\alpha$ and $^{208}$Pb; (b) the dashed line is M3Y total potential and dotted line is the total potential with added kinetic energy. The solid horizontal line indicates the $Q_{\alpha}$-value for the $^{212}Po$.}
\end{figure}

The calculated kinetic energy typically for $\alpha$ + $^{208}$Pb is presented by red dotted line in Fig.\ref{Fig02}(a). This figure indicates that the kinetic energy is impressive in the interior region, where two nuclei have sensible overlap. Also, its value is confronted by a gradual depression concerning the low overlapping densities at nuclear surfaces.

The estimated kinetic energy by the ETF approach for the dinuclear system can be accompanied as a corrective term with the DF model that explicitly presented in Fig.\ref{Fig02}(a). The presented results in Fig.\ref{Fig02}(b) indicate that the influence of the considered kinetic energy on the interior regions of the Coulomb barrier is quite evident. This kinetic energy due to the PEP acts as a repulsive force that hinders a large density overlap in the dinuclear system. Consequently, the half-life calculations would be affected by this modification. For seeking such influences, by employing the M3Y potentials and modified potentials with added kinetic energy terms, the half-lives of the even-even nuclei with $84\leq Z\leq92$ are calculated through WKB approximation that are listed as ${T^{(2)}_{1/2}}$ and ${T^{(3)}_{1/2}}$ in Tab.\ref{Tab.1}. The standard deviation (SD) of the 48 nuclei, $\sigma=\sqrt{\frac{1}{N}\sum_{i=1}^{N=48}[\log(T^{Calc}_{i}/T^{Exp}_{i})]^{2}}$, of the calculated half-lives from their corresponded experimental data is employed for having a better insight into this modification. The obtained SD value by use of WKB approximation for the standard M3Y potentials is $1.245$, which varies to $0.787$ for the half-lives obtained by modified potentials within kinetic energy considerations. This SD promotion indicates how the kinetic energy application to the DF model would be productive on the calculation of the $\alpha$ decay half-lives.

On the other hand, for considering the PEP in the total system, we renormalized the NN interactions according to the BS quantization condition. Consequently, the $\alpha$ decay half-lives are being calculated within the Gurvitz method, and their logarithms are expressed as ${T^{(1)}_{1/2}}$ in Tab.\ref{Tab.1}. The SD value 0.994 is obtained for this case. The obtained results express that the $\alpha$ decay half-lives that are being calculated by kinetic energy application to the DF model are more consistent with the experimental data in comparison with their corresponding values obtained by the renormalizing and Gurvitz method.

It is noticeable that all the $\alpha$ decay half-lives are calculated with $P_{\alpha}=1$ assumption. Extensively, it has shown that the alpha formation probability has a remarkable role in the $\alpha$ decay studies \cite{61,62,63}. However, if the calculations of the $\alpha$ decay half-lives within WKB approximation, in the PCM viewpoint, associates with the cluster preformation factors estimating by the CFM \cite{45,46,68} can result in a good agreement with the experimental data. The preformation factors calculated by CFM are presented as $P^{CFM}_{\alpha}$ in Tab.\ref{Tab.1}. The SD values for half-lives obtained by M3Y potentials change to $0.561$, and it reduces to $0.309$ for those obtained by modified potentials as a consequence of the application of the preformation factors estimated by the CFM. Also, the logarithms of calculated half-lives with $P_{\alpha}$ applications for M3Y and modified potentials with kinetic energies are presented as ${T^{(4)}_{1/2}}$ and ${T^{(5)}_{1/2}}$ in Tab.\ref{Tab.1}. Furthermore, if the alpha cluster preformation factors estimating by the CFM are being applied to half-lives calculated within the Gurvitz method, the SD value changes from 0.994 to 0.340.

Moreover, we employed some various phenomenological formulas based on the Geiger Nuttall law \cite{47} that have been proposed to estimate the $\alpha$ decay half-lives for comparing with the estimated half-lives. The Viola-Seaborg-Sobiczewski ($VSS$) semi-empirical relationship was determined as
\begin{equation}\label{Eq.24}
\log(T_{1/2})=\frac{(aZ+b)}{\sqrt{Q_{\alpha}}}+cZ+d+e
\end{equation}
where Z is the atomic number of the parent nucleus and the constants a, b, c, and d are $1.66175$, $-8.5166$, $-0.20228$, and $-33.9069$, respectively \cite{48}.
For the even-even nuclei the parameter $e$ is assumed to be zero \cite{55}.

Also, the Royer analytic formula ($RF$) for the $\alpha$ decay half-lives was proposed as \cite{49},
\begin{equation}\label{Eq.25}
\log(T_{1/2})=a+bA^{\frac{1}{6}}\sqrt{Z}+\frac{cZ}{\sqrt{Q_{\alpha}}}
\end{equation}
where A and Z denote the mass and atomic numbers of the parent nuclei. For the even-even nuclei the constants a, b, and c are $-25.31$, $-1.1629$, and $1.5864$, respectively.

Furthermore, the universal decay law ($UDL$) in charged-particle emission and exotic cluster radioactivity was presented by Qi et al.\cite{50}
\begin{equation}\label{Eq.26}
\log(T_{1/2})=aZ_{\alpha}Z_{d}\sqrt{\frac{A}{Q_{\alpha}}}+b\sqrt{AZ_{\alpha}Z_{d}(A_{\alpha}^{\frac{1}{3}}+A_{d}^{\frac{1}{3}})}+c
\end{equation}
here, $A=(A_{\alpha}A_{d})/(A_{\alpha}+A_{d})$ and constants $a=0.4314$, $b=-0.4087$, and $c=-25.7725$ were obtained by fitting to experimental of both alpha and cluster decays \cite{50}. The logarithms of the calculated half-lives obtained by VSS, RF, and UDL formulas are presented as ${T^{VSS}_{1/2}}$, ${T^{RF}_{1/2}}$, and ${T^{UDL}_{1/2}}$ in Tab.\ref{Tab.1}, respectively. The SD values of the calculated half-lives obtained by VSS, RF, and UDL formulas are $0.704$, $0.385$, and $0.484$, respectively. Also, by comparison on SD values of the calculated half-lives obtained by these mentioned phenomenological formulas and those obtained by M3Y and modified DF potentials, one would be clarified that the calculated half-lives within simultaneous kinetic energy modification and alpha preformation factor applications obtained by CFM along with the RF formula are more consistent with the experimental data. Therefore, the role of this modification in the $\alpha$ decay studies would be more impressive.

\begin{figure}
\centering
\includegraphics[width=4.5in,height=4.5in]{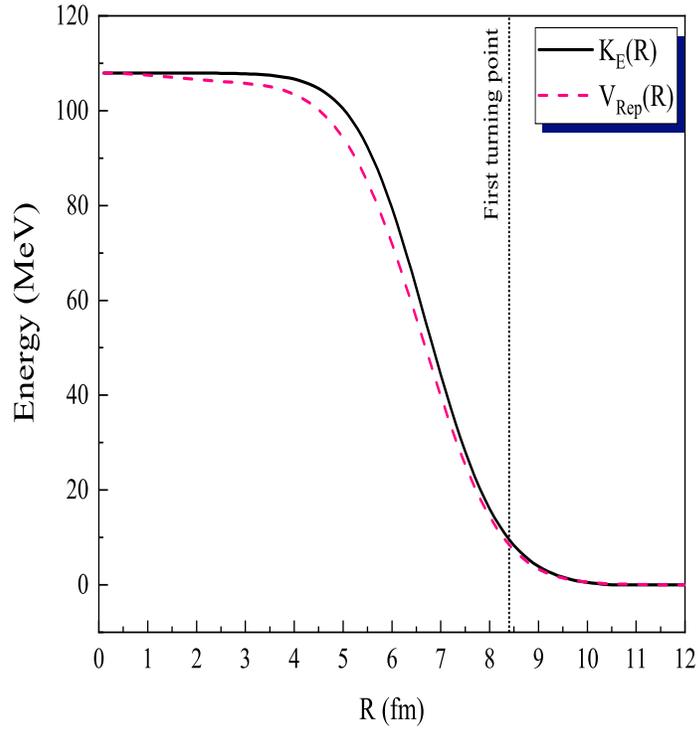}
\caption{\label{Fig03} Solid line is the kinetic energy estimated by ETF approach and dashed line is the simulated double-folded integral with a delta function, typically for the overlapping of $\alpha$ and $^{208}$Pb.}
\end{figure}

In order to simplify the estimation of kinetic energy through the sophisticated Hartree-Fock and ETF approaches, a double-folded integral with a delta function $V_{0}\delta(\textbf{s})$ is simulated as
\begin{equation}\label{Eq.27}
V_{Rep}(R)=V_{0}\iint \rho_{1}(\textbf{r}_{1}) \delta(\textbf{s})\rho_{2}(\textbf{r}_{2})\, d\textbf{r}_{1}d\textbf{r}_{2}.
\end{equation}
here, $V_{0}$ is the strength of this simulated repulsive force obtaining at where the alpha and daughter have a total overlap $(R=0)$. Its value is being adjusted so that to reproduce the estimated kinetic energy by the ETF approach, especially where the PEP becomes more apparent in the total system. For instance, the simulated repulsive force for the dinuclear system ($\alpha$ + $^{208}$Pb) is displayed in Fig.\ref{Fig03}. The result in this figure indicates that the overlapping of the $\alpha$ and daughter nuclei would be associated with some energy variations at nuclear surfaces. Furthermore, the presented results in this figure indicate that the treatment of our simulation is close to the treatment of kinetic energy with fair approximation. Also, the treatments and the magnitudes of such simulation and estimated kinetic energy are exciting the same from the first turning point and thereafter. Hence, this similarity at $R=0$ and from the first turning point is supporting the choice of such simulated double-folded integral with a delta function.

\begin{figure}
\centering
\includegraphics[width=4.5in,height=4.5in]{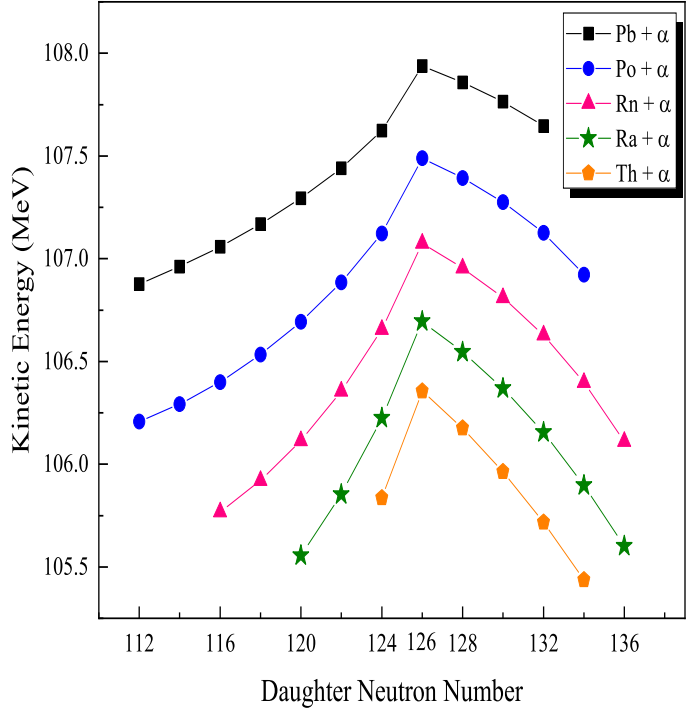}
\caption{\label{Fig04} The kinetic energy values for the overleaping $\alpha$ and daughter nuclei examined at $R=0$ with respect to daughter neutron number.}
\end{figure}

Concerning to the calculated kinetic energy for selected isotopic groups, their values at origin are being determined and presented in Fig.\ref{Fig04}. The results in Fig.\ref{Fig04} indicate the quantized shifts for examined kinetic energies according to an increase in atomic numbers. Also, for the specific isotope group, the kinetic energy has a peak for the parent nucleus that its daughter has a magic neutron closed-shell, which can be as a result of the shell-closure effect at magic neutron number $N = 126$. One can expect that the PEP causes a nucleon interacting space at full strength. It more increases the kinetic energy during the density overlap of alpha and the nuclei with the full occupied layers, especially magic nuclei. This treatment can be justified as the fully occupied valence layer in the magic nuclei, causing a higher nucleon accumulation in constant volume in comparison with the nuclei that have a non-full valence layer.

It is noticeable that each nucleus has its characteristic $f_{q}(r)$ that relates to the density distributions of the nucleons calculated by the SLy4, in this paper.  For examining the $f_{q}(r)$ contributions to the shell-closure treatment displayed in Fig.\ref{Fig04}, their values and the amount of the nucleon densities for the $^{206-210}$Pb are being estimated at $R=0$, where to nuclei have complete overlap. The estimated effective mass form factor $f_{p}(r)$ for $^{206}$Pb, $^{208}$Pb, and $^{210}$Pb are 1.4344, 1.4366, and 1.4328, respectively. Also, the estimated effective mass form factor $f_{n}(r)$ for $^{206}$Pb, $^{208}$Pb, and $^{210}$Pb are 1.4595, 1.4628, and 1.4620, respectively. On the other hand, by considering the dependency of $f_{q}(r)$ to the variation of the total energy $E$ with respect to the kinetic energy density, $f_{q}(r)=(2M^{q}/\hbar^{2})(\delta E/\delta\tau$), one can expect that the higher effective mass form factor would result in a higher kinetic energy. According to the estimated $f_{p}(r)$ and $f_{n}(r)$, the double closed-shell nucleus like $^{208}$Pb has more kinetic energy than the isotopes in its vicinity, as indicated in Fig.\ref{Fig04}. Hence one can expect that the shell-closure treatment in Fig.\ref{Fig04} can be due to the effective mass form factors that are being obtained by density distributions calculated by SLy4 force.

\begin{figure}
\centering
\includegraphics[width=4.0in,height=4.5in]{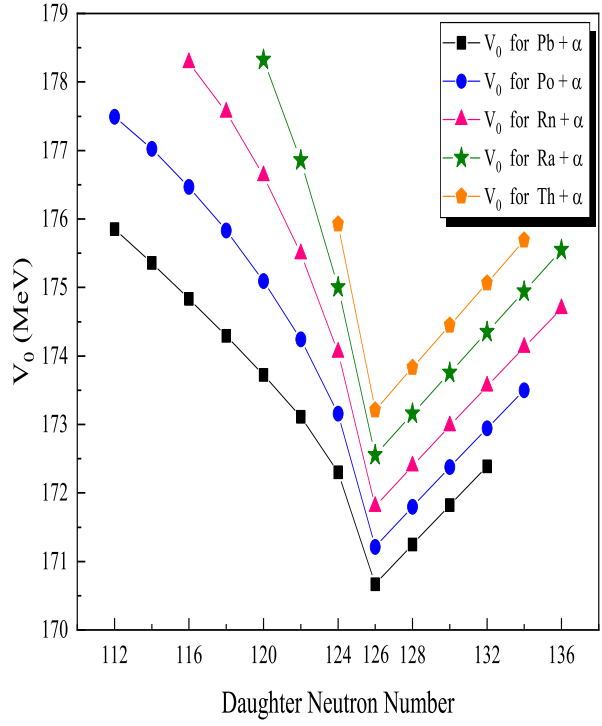}
\caption{\label{Fig05} The strength of the simulated double-folded integral with a delta function for the overleaping $\alpha$ and daughter nuclei adjusted at $R = 0$ due to the kinetic energy.}
\end{figure}

The all adjusted $V_0$ values at $R = 0$ with respect to the daughter neutron numbers for the investigated heavy nuclei are presented in Fig.\ref{Fig05}. The results in this figure indicate an obvious shell-closure effect for a characterized isotope group. As shown in Fig.\ref{Fig05}, quadratic dependencies with respect to the daughter neutron number are obtained for the nuclei with valence holes and linear dependencies for the isotopes with the valence particles.

Instead of discrete calculation for the strength of the such simulation, the mentioned two different dependencies made us drawn to fit these obtained $V_0$ values with respect to the atomic and mass numbers of daughter nuclei. These fitted relations are illustrated as below:
\begin{equation}\label{Eq.28}
  V_{0} =\begin{cases}
    a+bZ_{d}+(cZ_{d}+d)A_{d}^{1/3}\sqrt{N_{0}-N_{d}}, & \text{$N_{d}<N_{0}$},\\
    a+b(A_{d}-N_{0}). & \text{$N_{d}\geq N_{0}$.}
  \end{cases}
\end{equation}
where the coefficients $a = 145$, $b = 0.3125$, $c = 0.0255$, and $d = 1.8575$ are obtained and $N_{0}$ is equivalent to the neutron magic number $N=126$. $A_{d}$, $Z_d$, and $N_d$ are the mass, atomic, and neutron numbers of daughter nuclei, respectively. Also, our simulation can compensate for the non-consideration of the kinetic energy in the DF potential as like as the estimated kinetic energy through the extended Thomas-Fermi approach. The $\alpha$ decay half-lives are being calculated upon implementing the simulated double-folded integral with a delta function and $P_{\alpha}$ application obtained by CFM. Their logarithms are presented as ${T^{(6)}_{1/2}}$ in Tab.\ref{Tab.1}. The obtained SD for the calculated half-lives through the modified DF potentials with this simulation is $0.298$, which indicates a good consistency with SD values obtained by RF formula and those obtained by the DF potential with the kinetic energy modification and $P_{\alpha}$ application. Although both kinetic energy consideration and the NN renormalization due to the BS quantization condition in the DF model can reproduce the effect of the PEP in the total system, the calculated $\alpha$ decay half-lives within the kinetic energy consideration are more consistent with the experimental data. The detailed properties of all considered states are presented in Tab.\ref{Tab.2}.

\section{\label{sec.4}SUMMARY AND CONCLUSION}

In this study, the effects of the repulsive forces arising from the PEP are investigated on the calculations of the $\alpha$ decay half-life for even-even nuclei with $84\leq Z\leq92$. Such repulsive force between NN interactions is investigated by considering an increase in the kinetic energy in the dinuclear systems. The densities of the alpha and daughter nuclei have an overlap. Also, the effect of the PEP is investigated on the $\alpha$ decay by the renormalizing the NN interactions by the BS quantization condition. To this intention, the ETF approach is used to estimate such kinetic energies that have arisen at the overlapping regions. Subsequently, the results expose some energy variation at nuclear surfaces causes by the applied interior modifications to the DF formalism, which can affect the $\alpha$ decay half-life calculations.

The SD value for the half-lives obtained by the standard M3Y potentials is 1.245, which reduces to 0.787 for modified M3Y potentials by the PEP inclusion by increasing kinetic energy consideration at overlapping regions. These values varied to 0.561 and 0.309 by the $P_{\alpha}$ application on the half-lives obtained by M3Y and modified M3Y potentials, respectively. On the other hand, the SD value 0.340 is being obtained for the normalized potentials within the BS condition for embedding the PEP. These obtained results reveal that the $\alpha$ decay half-lives would be more affected by the repulsive force considerations on the DF formalism, which proceed to the more consistent half-lives with the experimental data. Moreover, we proposed a pocket formula instead of the kinetic energy estimations through the sophisticated Hartree-Fock and ETF approaches, associated with DF formalism.

\clearpage
\clearpage

\clearpage

\clearpage

\clearpage

\begin{table*}
\caption{\label{Tab.1} A comparison between the logarithms of the calculated $\alpha$ decay half-lives and $T_{1/2}$(VSS), $T_{1/2}$(RF), and $T_{1/2}$(UDL). The half-lives are being calculated in the unit of sec.}
\begin{ruledtabular}
\begin{tabular}{ccccccccccccc}
 ${Parent}$ & ${Q_{\alpha}}$[MeV] & ${T^{Exp}_{1/2}}$ & $P_{\alpha}^{CFM}$ &${T^{(1)}_{1/2}}$ &${T^{(2)}_{1/2}}$ &${T^{(3)}_{1/2}}$& ${T^{(4)}_{1/2}}$ &  ${T^{(5)}_{1/2}}$ &  ${T^{(6)}_{1/2}}$ &${T^{VSS}_{1/2}}$& ${T^{RF}_{1/2}}$ & ${T^{UDL}_{1/2}}$ \\
 \hline\\
   $^{198}_{84}$Po   &  6.3096  &  2.2678  &    0.206  &    1.2479  &  0.9071  &  1.3917  &   1.5932  &   2.0778  &  2.0677  &   1.2815  &   2.0095  &   1.8459   \\[2pt]
   $^{200}_{84}$Po   &  5.9814  &  3.7939  &    0.188  &    2.6637  &  2.3218  &  2.8150  &   3.0476  &   3.5408  &  3.5264  &   2.6939  &   3.4024  &   3.3204   \\[2pt]
   $^{202}_{84}$Po   &  5.7010  &  5.1442  &    0.174  &    3.9741  &  3.6300  &  4.1316  &   4.3895  &   4.8911  &  4.8715  &   3.9961  &   4.6834  &   4.6776   \\[2pt]
   $^{204}_{84}$Po   &  5.4848  &  6.2768  &    0.158  &    5.0466  &  4.7021  &  5.2120  &   5.5034  &   6.0133  &  5.9863  &   5.0675  &   5.7304  &   5.7887   \\[2pt]
   $^{206}_{84}$Po   &  5.3269  &  7.1446  &    0.146  &    5.8648  &  5.5172  &  6.0368  &   6.3528  &   6.8724  &  6.8332  &   5.8909  &   6.5254  &   6.6352   \\[2pt]
   $^{208}_{84}$Po   &  5.2153  &  7.9612  &    0.135  &    6.3954  &  6.0737  &  6.6333  &   6.9434  &   7.5030  &  7.4340  &   6.4953  &   7.0982  &   7.2478   \\[2pt]
   $^{210}_{84}$Po   &  5.4074  &  7.0776  &    0.104  &    5.3241  &  4.9776  &  5.4980  &   5.9606  &   6.4810  &  6.4409  &   5.4666  &   6.0109  &   6.1146   \\[2pt]
   $^{212}_{84}$Po   &  8.9541  & -6.5243  &    0.220  &   -7.6543  & -7.9059  & -7.4644  &  -7.2483  &  -6.8068  & -6.8842  &  -7.0964  &  -6.8029  &  -7.3526   \\[2pt]
   $^{214}_{84}$Po   &  7.8334  & -3.7844  &    0.213  &   -4.6055  & -4.9360  & -4.4335  &  -4.2644  &  -3.7619  & -3.8385  &  -4.0678  &  -3.7645  &  -4.1478   \\[2pt]
   $^{216}_{84}$Po   &  6.9063  & -0.8386  &    0.206  &   -1.6597  & -1.9110  & -1.3640  &  -1.2249  &  -0.6779  & -0.7697  &  -1.0235  &  -0.7098  &  -0.9255   \\[2pt]
   $^{218}_{84}$Po   &  6.1146  &  2.2696  &    0.196  &    1.5048  &  1.2416  &  1.8188  &   1.9493  &   2.5265  &  2.4273  &   2.1070  &   2.4328  &   2.3898   \\[2pt]
   $^{200}_{86}$Rn   &  7.0433  &  0.0783  &    0.229  &   -0.7716  & -1.0720  & -0.6157  &  -0.4318  &   0.0245  &  0.0343  &  -0.6632  &   0.0177  &  -0.2277   \\[2pt]
   $^{202}_{86}$Rn   &  6.7737  &  1.0947  &    0.212  &    0.2141  & -0.0878  &  0.3878  &   0.5859  &   1.0615  &  1.0712  &   0.3346  &   0.9875  &   0.8025   \\[2pt]
   $^{204}_{86}$Rn   &  6.5464  &  2.0118  &    0.195  &    1.0671  &  0.7828  &  1.2500  &   1.4928  &   1.9600  &  1.9612  &   1.2235  &   1.8468  &   1.7166   \\[2pt]
   $^{206}_{86}$Rn   &  6.3838  &  2.7393  &    0.178  &    1.6531  &  1.3878  &  1.9021  &   2.1374  &   2.6517  &  2.6223  &   1.8882  &   2.4791  &   2.3919   \\[2pt]
   $^{208}_{86}$Rn   &  6.2606  &  3.3723  &    0.162  &    2.2573  &  1.8872  &  2.3763  &   2.6777  &   3.1668  &  3.1638  &   2.4090  &   2.9655  &   2.9138   \\[2pt]
   $^{210}_{86}$Rn   &  6.1589  &  3.9542  &    0.152  &    2.3159  &  2.3063  &  2.8031  &   3.1245  &   3.6213  &  3.6075  &   2.8507  &   3.3720  &   3.3515   \\[2pt]
   $^{212}_{86}$Rn   &  6.3850  &  3.1565  &    0.120  &    1.6317  &  1.2782  &  1.7801  &   2.1990  &   2.7009  &  2.6409  &   1.8832  &   2.3483  &   2.2841   \\[2pt]
   $^{214}_{86}$Rn   &  9.2084  & -6.5686  &    0.228  &   -7.6792  & -7.8166  & -7.3759  &  -7.1745  &  -6.7338  & -6.8006  &  -7.0148  &  -6.7258  &  -7.2564   \\[2pt]
   $^{216}_{86}$Rn   &  8.1973  & -4.3468  &    0.236  &   -4.9573  & -5.2271  & -4.7531  &  -4.6000  &  -4.1260  & -4.1687  &  -4.3628  &  -4.0746  &  -4.4570   \\[2pt]
   $^{218}_{86}$Rn   &  7.2625  & -1.4559  &    0.234  &   -1.9789  & -2.3232  & -1.7862  &  -1.6924  &  -1.1554  & -1.2314  &  -1.4333  &  -1.1412  &  -1.3603   \\[2pt]
   $^{220}_{86}$Rn   &  6.4046  &  1.7451  &    0.220  &    1.1712  &  0.9214  &  1.4912  &   1.5790  &   2.1488  &  2.0643  &   1.8017  &   2.1025  &   2.0636   \\[2pt]
   $^{222}_{86}$Rn   &  5.5903  &  5.5186  &    0.221  &    4.8612  &  4.7161  &  5.3124  &   5.3717  &   5.9680  &  5.8786  &   5.5380  &   5.8554  &   6.0240   \\[2pt]
   $^{206}_{88}$Ra   &  7.4151  & -0.6198  &    0.222  &   -1.3163  & -1.6664  & -1.2171  &  -1.0128  &  -0.5635  & -0.5391  &  -1.1331  &  -0.5542  &  -0.7904   \\[2pt]
   $^{208}_{88}$Ra   &  7.2730  &  0.1361  &    0.201  &   -0.8534  & -1.1977  & -0.7436  &  -0.5009  &  -0.0468  & -0.0216  &  -0.6415  &  -0.0985  &  -0.3008   \\[2pt]
   $^{210}_{88}$Ra   &  7.1508  &  0.5859  &    0.184  &   -0.4438  & -0.7897  & -0.3332  &  -0.0545  &   0.4020  &  0.4164  &  -0.2070  &   0.2994  &   0.1280   \\[2pt]
   $^{212}_{88}$Ra   &  7.0316  &  1.1845  &    0.170  &   -0.0321  & -0.3890  &  0.0858  &   0.3806  &   0.8554  &  0.8251  &   0.2276  &   0.6980  &   0.5574   \\[2pt]
   $^{214}_{88}$Ra   &  7.2725  &  0.3912  &    0.139  &   -0.9566  & -1.3345  & -0.8564  &  -0.4775  &   0.0006  & -0.0078  &  -0.6397  &  -0.2229  &  -0.4027   \\[2pt]
   $^{216}_{88}$Ra   &  9.5257  & -6.7399  &    0.239  &   -7.7321  & -7.8831  & -7.5032  &  -7.2615  &  -6.8816  & -6.8251  &  -7.0864  &  -6.7992  &  -7.3190   \\[2pt]
   $^{218}_{88}$Ra   &  8.5459  & -4.5986  &    0.241  &   -5.4512  & -5.4652  & -4.9964  &  -4.8472  &  -4.3784  & -4.4146  &  -4.5978  &  -4.3177  &  -4.6967   \\[2pt]
   $^{220}_{88}$Ra   &  7.5924  & -1.7447  &    0.239  &   -2.3569  & -2.6258  & -2.1293  &  -2.0042  &  -1.5077  & -1.5406  &  -1.7271  &  -1.4485  &  -1.6656   \\[2pt]
   $^{222}_{88}$Ra   &  6.6788  &  1.5798  &    0.199  &    0.9979  &  0.7055  &  1.2530  &   1.4066  &   1.9541  &  1.8845  &   1.5817  &   1.8652  &   1.8341   \\[2pt]
   $^{224}_{88}$Ra   &  5.7888  &  5.4964  &    0.183  &    4.9381  &  4.6929  &  5.2814  &   5.4304  &   6.0189  &  5.9432  &   5.5317  &   5.8292  &   6.0194   \\[2pt]
   $^{226}_{88}$Ra   &  4.8706  & 10.7029  &    0.181  &    9.5268  &  9.9942  & 10.6077  &  10.7365  &  11.3500  & 11.3099  &  10.6943  &  11.0226  &  11.5005   \\[2pt]
   $^{212}_{90}$Th   &  7.9579  & -1.4989  &    0.207  &   -2.4234  & -2.7755  & -2.3261  &  -2.0915  &  -1.6421  & -1.6094  &  -2.1148  &  -1.6370  &  -1.8932   \\[2pt]
   $^{214}_{90}$Th   &  7.8271  & -1.0605  &    0.193  &   -1.9449  & -2.3820  & -1.9234  &  -1.6676  &  -1.2090  & -1.1947  &  -1.6988  &  -1.2580  &  -1.4844   \\[2pt]
   $^{216}_{90}$Th   &  8.0723  & -1.5850  &    0.158  &   -2.7989  & -3.2102  & -2.7406  &  -2.4089  &  -1.9393  & -1.9431  &  -2.4704  &  -2.0809  &  -2.3421   \\[2pt]
   $^{218}_{90}$Th   &  9.8490  & -6.9318  &    0.252  &   -7.9102  & -7.9604  & -7.5285  &  -7.3618  &  -6.9299  & -6.9776  &  -7.1704  &  -6.8803  &  -7.3902   \\[2pt]
   $^{220}_{90}$Th   &  8.9530  & -5.0132  &    0.247  &   -5.6070  & -5.8344  & -5.3725  &  -5.2271  &  -4.7652  & -4.8237  &  -4.9752  &  -4.6993  &  -5.0830   \\[2pt]
   $^{222}_{90}$Th   &  8.1269  & -2.6498  &    0.231  &   -3.3062  & -3.5433  & -3.0574  &  -2.9069  &  -2.4210  & -2.4496  &  -2.6374  &  -2.3737  &  -2.6231   \\[2pt]
   $^{224}_{90}$Th   &  7.2985  &  0.1461  &    0.198  &   -0.5775  & -0.8251  & -0.3186  &  -0.1218  &   0.3847  &  0.3589  &   0.0948  &   0.3515  &   0.2584   \\[2pt]
   $^{226}_{90}$Th   &  6.4508  &  3.2634  &    0.182  &    2.5613  &  2.5199  &  3.0792  &   3.2598  &   3.8191  &  3.7646  &   3.4192  &   3.6765  &   3.7724   \\[2pt]
   $^{228}_{90}$Th   &  5.5200  &  7.7798  &    0.183  &    7.1956  &  7.1081  &  7.6960  &   7.8456  &   8.4335  &  8.3742  &   7.9188  &   8.1915  &   8.5414   \\[2pt]
   $^{218}_{92}$U    &  8.7747  & -3.2924  &    0.186  &   -4.2692  & -4.4752  & -4.0493  &  -3.7447  &  -3.3188  & -3.3451  &  -3.7813  &  -3.4037  &  -3.7199   \\[2pt]
   $^{222}_{92}$U    &  9.4302  & -5.3279  &    0.248  &   -6.1426  & -6.3565  & -5.9026  &  -5.7510  &  -5.2971  & -5.3416  &  -5.5056  &  -5.2300  &  -5.6274   \\[2pt]
   $^{224}_{92}$U    &  8.6198  & -3.0757  &    0.229  &   -4.0262  & -4.2472  & -3.7702  &  -3.6070  &  -3.1300  & -3.1537  &  -3.3453  &  -3.0871  &  -3.3588   \\[2pt]
   $^{226}_{92}$U    &  7.7009  & -0.5719  &    0.206  &   -1.1909  & -1.4206  & -0.9212  &  -0.7345  &  -0.2351  & -0.2506  &  -0.4943  &  -0.2456  &  -0.3531   \\[2pt]
   $^{228}_{92}$U    &  6.8035  &  2.7595  &    0.188  &    2.1752  &  1.9202  &  2.4387  &   2.6460  &   3.1645  &  3.1530  &   2.8303  &   3.0751  &   3.1583   \\[3pt]

\end{tabular}
\end{ruledtabular}
\end{table*}

\begin{table*}
\caption{\label{Tab.2} The detailed information of each states that are being investigated.}
\begin{ruledtabular}
\begin{tabular}{cccccc}
Potential & Half-life calculation method & $\lambda$ & $P_{\alpha}$ & SD & Top name in Tab.\ref{Tab.1}.\\
 \hline\\
  M3Y                &  Gurvitz \& Wildermuth condition                   &    $\lambda\neq1$  &   $P_{\alpha}=1$    &  0.994 &  ${T^{(1)}_{1/2}}$   \\[5pt]
  M3Y                &  Gurvitz \& Wildermuth condition \& $P_{\alpha}$   &    $\lambda\neq1$  &   $P_{\alpha}=CFM$  &  0.340 &        ---           \\[5pt]
  M3Y                &  WKB                                               &    $\lambda=1$     &   $P_{\alpha}=1$    &  1.245 &  ${T^{(2)}_{1/2}}$   \\[5pt]
  M3Y+Kinetic energy &  WKB                                               &    $\lambda=1$     &   $P_{\alpha}=1$    &  0.787 &  ${T^{(3)}_{1/2}}$   \\[5pt]
  M3Y                &  WKB \& $P_{\alpha}$                               &    $\lambda=1$     &   $P_{\alpha}=CFM$  &  0.561 &  ${T^{(4)}_{1/2}}$   \\[5pt]
  M3Y+Kinetic energy &  WKB \& $P_{\alpha}$                               &    $\lambda=1$     &   $P_{\alpha}=CFM$  &  0.309 &  ${T^{(5)}_{1/2}}$   \\[5pt]
  M3Y+Simulated term &  WKB \& $P_{\alpha}$                               &    $\lambda=1$     &   $P_{\alpha}=CFM$  &  0.298 &  ${T^{(6)}_{1/2}}$   \\[5pt]

\end{tabular}
\end{ruledtabular}
\end{table*}

\end{document}